\documentclass[aps,pra,twocolumn,amsmath,amssymb,superscriptaddress,linenumbers]{revtex4-1}

\usepackage{graphicx}
\usepackage{bm}
\usepackage{amsmath}
\usepackage{verbatim}     
\usepackage[dvipsnames]{xcolor}       
\usepackage[normalem]{ulem}   

\newcommand{\ket}[1]{|#1\rangle}                      		
\newcommand{\bra}[1]{\langle #1|}                     		
\newcommand{\beq}{\begin{equation}}
\newcommand{\eeq}{\end{equation}}
\newcommand{\bei}{\begin{itemize}}			
\newcommand{\eei}{\end{itemize}}			
\newcommand{\bc}{\mathbf{c}}
\newcommand{\bn}{\mathbf{n}}

\newcommand{\vect}[1]{{\mathbf #1}}
\newcommand{\vectgr}[1]{{\boldsymbol#1}}    		
\newcommand{\chiru}{\ket{\!\!\uparrow}}    			
\newcommand{\chird}{\ket{\!\!\downarrow}}    		

\newcommand{\cd}{\mathcal{C}}    

\usepackage{hyperref}
\hypersetup{%
   pdfpagemode=None, 
   pdfstartpage=1,
   pdfmenubar=true,
   pdftoolbar=true,
   colorlinks = true,
   linkcolor=blue,
   citecolor=blue,
   urlcolor=blue,
   bookmarksopen=false
 }

\begin{document}
\nolinenumbers
\title{Detection of Zak phases and topological invariants in a chiral quantum walk of twisted photons}

\author{Filippo Cardano}\email{filippo.cardano2@unina.it}
\affiliation{Dipartimento di Fisica, Universit\`{a} di Napoli Federico II, Complesso Universitario di Monte Sant'Angelo, Via Cintia, 80126 Napoli, Italy}
\author{Alessio D'Errico}
\affiliation{Dipartimento di Fisica, Universit\`{a} di Napoli Federico II, Complesso Universitario di Monte Sant'Angelo, Via Cintia, 80126 Napoli, Italy}
\author{Alexandre Dauphin}\email{alexandre.dauphin@icfo.es}
\affiliation{ICFO-Institut de Ciencies Fotoniques, The Barcelona Institute of Science and Technology, Av.\ Carl Friedrich Gauss 3, 08860 Castelldefels, Spain}
\author{Maria Maffei}
\affiliation{Dipartimento di Fisica, Universit\`{a} di Napoli Federico II, Complesso Universitario di Monte Sant'Angelo, Via Cintia, 80126 Napoli, Italy}
\affiliation{ICFO-Institut de Ciencies Fotoniques, The Barcelona Institute of Science and Technology, Av.\ Carl Friedrich Gauss 3, 08860 Castelldefels, Spain}
\author{Bruno Piccirillo}
\affiliation{Dipartimento di Fisica, Universit\`{a} di Napoli Federico II, Complesso Universitario di Monte Sant'Angelo, Via Cintia, 80126 Napoli, Italy}
\author{Corrado de Lisio}
\affiliation{Dipartimento di Fisica, Universit\`{a} di Napoli Federico II, Complesso Universitario di Monte Sant'Angelo, Via Cintia, 80126 Napoli, Italy}
\affiliation{CNR-SPIN, Complesso Universitario di Monte Sant'Angelo, Via Cintia, 80126 Napoli, Italy}
\author{Giulio De Filippis}
\affiliation{Dipartimento di Fisica, Universit\`{a} di Napoli Federico II, Complesso Universitario di Monte Sant'Angelo, Via Cintia, 80126 Napoli, Italy}
\affiliation{CNR-SPIN, Complesso Universitario di Monte Sant'Angelo, Via Cintia, 80126 Napoli, Italy}
\author{Vittorio Cataudella}
\affiliation{Dipartimento di Fisica, Universit\`{a} di Napoli Federico II, Complesso Universitario di Monte Sant'Angelo, Via Cintia, 80126 Napoli, Italy}
\affiliation{CNR-SPIN, Complesso Universitario di Monte Sant'Angelo, Via Cintia, 80126 Napoli, Italy}
\author{Enrico Santamato}
\affiliation{Dipartimento di Fisica, Universit\`{a} di Napoli Federico II, Complesso Universitario di Monte Sant'Angelo, Via Cintia, 80126 Napoli, Italy}
\affiliation{CNR-SPIN, Complesso Universitario di Monte Sant'Angelo, Via Cintia, 80126 Napoli, Italy}
\author{Lorenzo Marrucci}
\affiliation{Dipartimento di Fisica, Universit\`{a} di Napoli Federico II, Complesso Universitario di Monte Sant'Angelo, Via Cintia, 80126 Napoli, Italy}
\affiliation{CNR-ISASI, Institute of Applied Science and Intelligent Systems, Via Campi Flegrei 34, 80078 Pozzuoli (NA), Italy}
\author{Maciej Lewenstein}
\affiliation{ICFO-Institut de Ciencies Fotoniques, The Barcelona Institute of Science and Technology, Av.\ Carl Friedrich Gauss 3, 08860 Castelldefels, Spain}
\affiliation{ICREA -- Instituci{\'o} Catalana de Recerca i Estudis Avan\c{c}ats, Pg.\ Lluis Companys 23, E-08010 Barcelona, Spain}
\author{Pietro Massignan}
\affiliation{ICFO-Institut de Ciencies Fotoniques, The Barcelona Institute of Science and Technology, Av.\ Carl Friedrich Gauss 3, 08860 Castelldefels, Spain}


\begin{abstract}
Topological insulators are fascinating states of matter exhibiting protected edge states and robust quantized features in their bulk. 
Here, we propose and validate experimentally a method to detect topological properties in the bulk of one-dimensional chiral systems. 
We first introduce the mean chiral displacement, an observable that rapidly approaches a value proportional to the Zak phase during the free evolution of the system.
Then we measure the Zak phase in a photonic quantum walk of twisted photons, by observing the mean chiral displacement in its bulk.
Next, we measure the Zak phase in an alternative, inequivalent timeframe, and combine the two windings to characterize the full phase diagram of this Floquet system. 
Finally, we prove the robustness of the measure by introducing dynamical disorder in the system. 
This detection method is extremely general, and readily applicable to all present one-dimensional platforms simulating static or Floquet chiral systems.
\end{abstract}

\maketitle

\begin{figure*}
\centering
\vskip 0 pt
\includegraphics[scale=1]{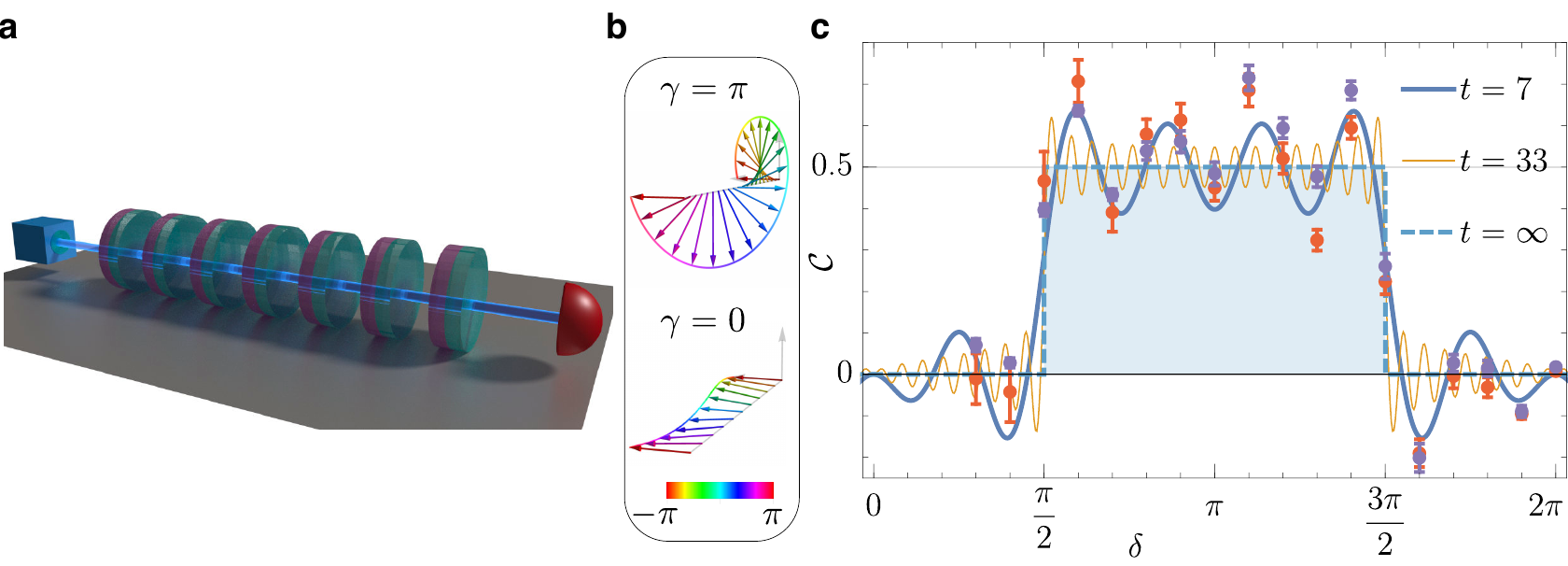}
\caption{{\bf Zak phase detection through the mean chiral displacement.} 
(a)  Sketch of the setup implementing the protocol $U=Q\cdot W$.
A light beam, exiting a single mode fiber depicted on the left, performs a QW by propagating through a sequence of quarter-wave plates (purple disks) and $q$-plates (turquoise disks). 
(b) The unit vector $\bn(k)$ winds either 1 or 0 times around the chiral axis, as $k$ traverses the whole Brillouin zone, depending on the value of the optical retardation $\delta$.
(c) Mean chiral displacement $\cd$ after a 7-steps QW of protocol $U$,  vs.\ the optical retardation $\delta$.
Each datapoint is an average over ten different measurements (error bars are the associated standard errors). Purple and red dots refer, respectively, to different input polarizations, $\ket{L}$ and $(\ket{L}+\ket{R})/\sqrt{2}$.
The lines represent the function $S_\Gamma(t)$ given in equation \eqref{SGamma}, for different values of the time  $t$. In the long time limit, $S_\Gamma(t)$ converges to (a multiple of) the Zak phase $\gamma$ of protocol $U$.
}
\label{fig:ZakPhaseOfU}
\end{figure*}

\section*{Introduction}
Topological phases of matter escape the canonical characterization of states dictated by the Ginzburg-Landau theory of phase transitions. These phases emerge without breaking symmetries and are not characterized by a long-range order nor a local order parameter, but rather by a global topological order.
Historically, topology was first proven to have a key role in explaining algebraically decaying  order,  transport and coherence of two-dimensional Bose liquids, XY-models and crystals \cite{Kosterlitz2016}.
Shortly after, the quantization of Hall conductance \cite{Klitzing1980} was shown to be rooted in current-carrying edge states, protected by the topology of the bulk \cite{Laughlin1981,Thouless1982,Halperin1982}.
Being associated to a global order, these phases are robust against local perturbations and promise important applications in metrology, spintronics, and quantum computation  (see, e.g.,
Refs.\ \cite{VonKlitzing1986,Fert2008,Pachos2012}).

Intense studies \cite{Qi2011} followed the early discoveries, and topological insulators have by now been engineered in a variety of physical architectures, such as superconducting \cite{Beenakker2016}, mechanical \cite{Huber2016}, optomechanical \cite{Peano2015},  photonic \cite{Lu2016}, atomic \cite{Goldman2016} and acoustic platforms \cite{Xiao2015}. Such diverse systems have been exposed to either real or synthetic magnetic fields, and their topological properties have been studied by scattering at the interface between different domains \cite{Xiao2015,Barkhofen2016}, or imaging edge states \cite{Kitagawa2012,Rechtsman2013,Hafezi2013,Stuhl2015,Mancini2015,Leder2016,Meier2016,Mukherjee2016,Peng2016,Maczewsky2017}. 
Direct detection of topological invariants in the bulk of the system (with no need of edges) has been reported so far by very few experiments \cite{Atala2013,Aidelsburger2015,Zeuner2015}.

Topological insulators are classified in terms of dimensionality and discrete symmetries \cite{Chyu2016}. One-dimensional (1D) systems with chiral symmetry are characterized by the Zak phase, i.e., the Berry phase accumulated by an eigenstate during its parallel transport through the whole Brillouin zone \cite{Zak1989}. The Zak phase is closely related to the electric polarization in solids and plays a key role in the modern theory of insulators \cite{Resta1994,Xiao2010}.

Periodically driven (Floquet) systems are attracting an increasing interest, as these show richer topological features than their static counterparts \cite{Oka2009,Lindner2011,Kitagawa2012,Asboth2012,Cayssol2013,Asboth2013, Rudner2013,Goldman2014,Asboth2014, Nathan2015,Mukherjee2016,Peng2016,Maczewsky2017,DalLago2015,Fleury2016}. 
Particularly promising Floquet topological systems are discrete-time Quantum Walks (QWs) \cite{Kitagawa2010a,Kitagawa2012,Zeuner2015,Barkhofen2016,Cardano2016,Groh2016}, 
and recent works have reported the observation of topological invariants \cite{Zeuner2015,Barkhofen2016}, quantum phase transitions \cite{Cardano2016} and edge states \cite{Kitagawa2012} in these systems.
In its simplest version, a QW is the discrete time evolution of a particle (the walker) on a 1D lattice \cite{Venegas-Andraca2012}. At each step, the walker moves to neighbouring sites, with the direction of the shift depending on the state of an internal two-level degree of freedom (the coin). Between consecutive steps, a rotation modifies the coin state, univoquely determining the following evolution. 

Here we demonstrate that, in chiral 1D static and Floquet systems with spin 1/2 (i.e., a two-state coin), the mean chiral displacement of a particle's wavepacket becomes quantized and proportional to the Zak phase in the long time limit.
Remarkably, this occurs during the free evolution of the system, in absence of any external force or loss mechanism, with the only requirement that the initial wavefunction is localized.
We validate experimentally this finding in a photonic discrete-time QW based on the orbital angular momentum of a light beam. We implement the same QW in a shifted inequivalent timeframe and measure a second Zak phase. 
Combining the two windings we extract the complete set of topological invariants characterizing the system. 
Finally, we prove the robustness of our detection by adding dynamical disorder. 

These measurements provide therefore a bulk measurement of the Zak phases and complete topological invariants of a 1D chiral quantum walk. Our proposal may be straightforwardly applied to general driven Floquet system.

\section*{Results}
{\bf Zak phase detection in the bulk of a quantum walk.}
In one dimension, discrete-time QWs with chiral symmetry display a quantized Zak phase and have been extensively studied in the past years. Among these implementations, we focus on the photonic platform proposed in Ref.\ \cite{Cardano2016}. Here, the walk takes place on a lattice whose sites $\ket{x}$ are associated with photonics states $\ket{m}$, corresponding to light beams that carry $m\hbar$ units of orbital angular momentum per photon along the propagation axis and show a twisted wavefront \cite{Yao11}.
The two coin states are instead mapped onto the left and right circular polarizations of the beam, carrying $\pm\hbar$ units of spin angular momentum per photon along the propagation axis.
Once the system is prepared in an initial state $\ket{\psi_0}$, its state after $t$ timesteps is given by
\beq
\ket{\psi(t)}=\mathcal{U}^t\ket{\psi_0},
\eeq
where the single-step operator $\mathcal{U}$ is obtained by cascading suitable combinations of quarter-wave plates and $q$-plates \cite{Marrucci2006,Cardano2015,Cardano2016}.
In Fig.\ \ref{fig:ZakPhaseOfU}a we show a pictorial representation of our setup that realizes a seven step quantum walk with $\mathcal{U}$ implemented specifically as $U\equiv Q \cdot W$ \cite{Cardano2016}.
The action of a quarter-wave plate oriented at $90^\circ$ with respect to the horizontal direction is described by the local operator $W$, rotating the polarization states as
\begin{align}\label{eq:qwp}
W=\frac{1}{\sqrt{2}}\sum_m \mathbf{c}^\dagger_m (\sigma_0-i\sigma_x) \mathbf{c}_{m}.
\end{align}
Here $\mathbf{c}^\dagger_m=(c^\dagger_{m,L},c^\dagger_{m,R})$ creates a particle on site $m$ with polarization L/R, and $\sigma_i$ are Pauli matrices acting in the coin (polarization) space.
The translation operator $Q$ is implemented by a $q$-plate, a liquid crystal device which yields an effective spin-orbit interaction in the light beam. This couples neighbouring sites and polarization states as 
\begin{align}\label{eq:qplate}
Q(\delta)=\sum_m \cos{\frac{\delta}{2}}\,\bc^\dagger_m \bc_m+i\sin{\frac{\delta}{2}}\left(\bc^\dagger_{m+1} \sigma_{-} \bc_m+{\rm h.c.} \right)
\end{align}
where $\sigma_\pm=(\sigma_x\pm i \sigma_y)/2$ are the operators that flip the coin states $\ket{L}$ and $\ket{R}$, $\delta$ is the optical retardation of the $q$-plate, {and h.c. stands for Hermitian conjugate}.
Further details on the $q$-plates and on the complete experimental setup are provided in the Methods and Supplementary Fig.\ 1.

\begin{figure*}[!ht]
\centering
\vskip 0 pt
\includegraphics[scale=1]{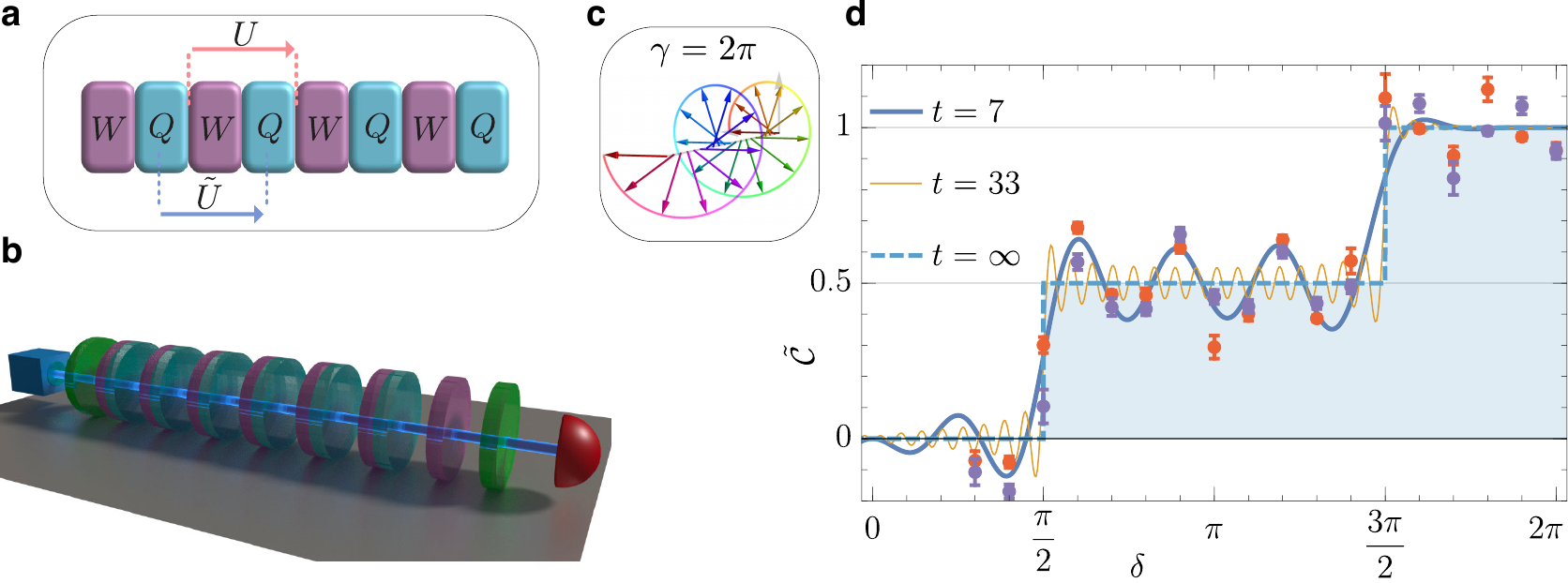}
\caption{{\bf Zak phase in the complementary timeframe.} 
(a) Different choices of the origin of the periodic cycle lead to different protocols.
(b) Sketch of the setup implementing protocol $\tilde U=\sqrt{Q}\cdot W \cdot \sqrt{Q}$. The two $q$-plates at the beginning and end of the optical path (shown in bright green) yield an optical retardation $\delta/2$, where $\delta$ is the optical retardation characterizing bulk $q$-plates (turquoise).
(c) The unit vector $\tilde \bn(k)$ associated to the operator $\tilde U$, for optical retardations $3\pi/2<\delta<2\pi$, winds twice around the chiral axis as $k$ spans the whole Brillouin zone. 
(d) Mean chiral displacement $\tilde{\cd}$ after a 7-steps QW  with protocol $\tilde U$.
The datapoints are averages of ten experimental measurements, and errorbars are the associated standard errors.
Purple and orange colors refer, respectively, to input polarizations $\ket{L}$ and $(\ket{L}+ i\,\ket{R})/\sqrt 2$. 
The lines display $S_{\tilde\Gamma}(t)$, obtained replacing ${\bf n}$ with $\tilde{\bf n}$ in equation \eqref{SGamma}, for different values of the time  $t$. At long times, $S_{\tilde\Gamma}$ converges to the Zak phase $\tilde\gamma$.
}
\label{fig:ZakPhaseOfUtilde}
\end{figure*}

Very generally, QWs are generated by the repeated application of a unitary operator $\mathcal{U}$, and therefore the system can be described in the framework of Floquet theory.
As a consequence of translational invariance in space, the effective Hamiltonian associated to a full period is diagonal in momentum {space} and may be written as 
\begin{align}
\mathcal{H}(k)=i \ln{\mathcal{U}}(k)=E(k) \bn(k)\cdot\vectgr{\sigma},
\end{align}
with $E(k)$ the quasi-energy dispersion, $\vectgr{\sigma}=(\sigma_x,\sigma_y,\sigma_z)$, and we have set the period $T$ and $\hbar$ to unity. The point on the Bloch sphere identified by the unit vector $\vect{n}(k)$ represents the coin part of the system eigenstates, while their spatial part is a plane wave with quasi-momentum $k$ \cite{Cardano2016}.
The function $\ln(x)$ denotes the principal branch of the natural (matrix) logarithm, so that the quasi-energy is a periodic function, with $-\pi$ and $+\pi$ identified.

The class of quantum walks we are considering features chiral symmetry, since there exists a unitary operator $\Gamma$ such that $\Gamma^2=I$, and $\Gamma \mathcal{H}\Gamma=-\mathcal{H}$. 
These conditions imply that $\Gamma$ is Hermitian and that $\Gamma=\vect{v}_\Gamma\cdot\vectgr{\sigma}$, with $\vect{v}_\Gamma$ a vector labeling a point on the Bloch sphere.
In this case, the unit vector $\vect{n}$ is bound to rotate around the origin in a plane orthogonal to $\vect{v}_\Gamma$, and the Zak phase equals
\beq
\gamma=\frac{1}{2}\int_{-\pi}^{\pi}{\rm d}k\;  
\left(\bn\times\frac{\partial \bn}{\partial k}\right)\cdot \vect{v}_\Gamma.
\eeq
The winding number $\gamma/\pi$ assumes strictly integer values and counts the number of times the unit vector $\bn$ rotates around the unit vector $\vect{v}_\Gamma$ as $k$ traverses the whole Brillouin zone. 
In Fig.~\ref{fig:ZakPhaseOfU}b we show the winding of the vector $\vect n$ of the operator $U$, for two values of $\delta$ in different topological sectors. 
The Zak phase is therefore a bulk property; although it has strong influences in  properties of systems where it arises, its detection in current experimental architectures remains challenging. 

In the following, we show that information on such topological invariant is hidden in the subleading terms of the mean displacement $\langle m \rangle$, when the initial wavepacket is localized on a single site. This extends a previous result showing that, in the same conditions, the ballistic terms of higher moments of the walker's displacement feature discontinuities at topological phase transitions \cite{Cardano2016}. Let us consider the evolution of a wavepacket $|\psi_0\rangle$ initially localized at site $m=0$, and whose polarization is characterized by  the expectation values of the three Pauli matrices, $\vect{s}=\langle \psi_0 | \vectgr{\sigma} | \psi_0\rangle=\langle \vectgr{\sigma} \rangle_{ \psi_0}$. 
The mean displacement of the wavepacket after $t$ timesteps is given by (see Supplementary Note 1 for details)

\begin{equation}
\label{meanDisplacement}
\begin{split}
\langle m(t)\rangle&= \int_{-\pi}^{\pi}\frac{{\rm d}k}{2\pi}\; \left\langle \mathcal{U}^{-t} {(i \partial_k)} \mathcal{U}^t \right\rangle_{\psi_0}\\
&= \langle \Gamma_\perp \rangle_{ \psi_0} [L(t)+S(t)] {-} \langle \Gamma \rangle_{ \psi_0} S_\Gamma(t).
\end{split}
\end{equation}
The term in square brackets in equation~\eqref{meanDisplacement} is proportional to $\langle \Gamma_\perp \rangle_{ \psi_0}$, the projection of the initial polarization on a direction orthogonal to $\vect v_\Gamma$, and contains a ballistic term $L(t)$ (which grows linearly with $t$) and a subleading part $S(t)$. 

The vector identifying the specific direction of $\Gamma_\perp$ in the plane orthogonal to $\vect v_\Gamma$,
 and the explicit functional forms of $L(t)$ and $S(t)$, are non-universal features which depend on the specific protocol (or timeframe), and have no particular relevance for our discussion. 
The second term in equation \eqref{meanDisplacement}, which is weighted by $\langle \Gamma \rangle_{ \psi_0}$ (the projection of the initial polarization along $\vect v_\Gamma$) is the subleading {chiral term} $S_\Gamma$, that may be written as (see Supplementary Note 1 for details)
\beq\label{SGamma}
S_\Gamma(t)=\frac{\gamma}{2\pi}
-\int_{-\pi}^{\pi}\frac{{\rm d}k}{2\pi}\;  \frac{\cos(2tE)}{2} 
\left(\vect{n}\times\frac{\partial \vect{n}}{\partial k}\right)\cdot \vect{v}_\Gamma.
\eeq
In the limit  $t\rightarrow\infty$, $S_\Gamma$ becomes proportional to the Zak phase, as the oscillatory correction quickly averages to zero (see Fig.\ \ref{fig:ZakPhaseOfU}c).

The above analysis shows that information on the Zak phase is contained in the mean displacement of the walker, and it may be extracted by fitting $\langle m\rangle$ at long times,  isolating in turn the second term of equation \eqref{meanDisplacement}.
A related result for the case of a non-Hermitian quantum walk initialized on a chiral eigenstate (i.e., an initial condition such that $\langle \Gamma_\perp \rangle_{ \psi_0}=0$) was demonstrated theoretically in Ref.\ \cite{Rudner2009}, and verified experimentally in Ref.\ \cite{Zeuner2015}.
However, this measurement would not be robust. Indeed, even if one prepared the initial polarization in an eigenstate of the chiral operator $\Gamma$, so that $\langle\Gamma_\perp \rangle_{ \psi_0}=0$, disorder during the propagation of the beam would introduce polarization components orthogonal to $\vect{v}_\Gamma$.
These would give rise to ballistic contributions, which in the long time limit would dramatically affect the result. 

An alternative and more convenient approach consists in measuring the mean chiral displacement
\beq\label{eq:meanChiralDisplacement}
\cd(t) \equiv 
\langle \Gamma m(t)\rangle = S_\Gamma(t),
\eeq
which quantifies the relative shift between the two projections of the state onto the eigenstates of the chiral operator (see Supplementary Note 1 for a concise derivation of this equality). Importantly, the result contained in equation \eqref{eq:meanChiralDisplacement} is (i) independent of the initial polarization and (ii) robust against disorder. 
We probe the chiral displacement in our photonic platform by performing a 7-step quantum walk of the protocol $U=Q\cdot W$, as depicted in Fig.\ \ref{fig:ZakPhaseOfU}a. The chiral eigenstates correspond to two specific orthogonal polarization states, which depend explicitly on the protocol, and which we detect at the end of the quantum walk (see Methods).
In Fig.\ \ref{fig:ZakPhaseOfU}c, we report the measured values of $\cd$ for two different initial polarization states. 
Experimental points closely follow the theory curve for 7 time steps (blue solid line), and no significant differences can be observed between the two different initial states, proving that this measurement is insensitive to the choice of the polarization of the photons. For completeness we also show results predicted for 33 steps, and the asymptotic long-time limit, which coincides with the Zak phase (over $2\pi$).  We note here that, although both theory and data oscillate, as few as 7 steps are enough to have a clear detection of the Zak phase.

{\bf Zak phase in a shifted timeframe.}
In static systems, bulk topological invariants such as the Zak phase or the Chern number are uniquely defined by integrals over the whole Brillouin zone, and are in one-to-one correspondence with the presence of edge states, thus providing a full classification in terms of the periodic table of topological insulators \cite{Chyu2016}.
The situation is  very different in periodically-driven (Floquet) systems in $D$ dimensions, where the integral determining the topological invariants needs to be performed over a $D+1$ dimensional torus constituted by the Brillouin zone and an extra periodic dimension, the quasi-energy \cite{Nathan2015}.

Moreover, a gauge freedom is introduced by the choice of the {
timeframe},
 i.e., the origin of time of the periodic cycle (see Fig.\ \ref{fig:ZakPhaseOfUtilde}a).
While the dispersion $E(k)$ is equal in all timeframes, the effective Hamiltonian, its eigenstates and symmetries, and the resulting dynamics crucially depend on the timeframe \cite{Goldman2014}.
As an example, the operator $\tilde{U} \equiv \sqrt{Q} \cdot W \cdot \sqrt{Q}$ defines a timeframe which is inequivalent to the one introduced by $U$. In particular, the unit vector $\tilde {\bf n}(k)$ defined by $i {\rm ln}\tilde U(k)=E(k) \tilde{\bf n}(k)\cdot\vectgr{\sigma}$ may wind twice around the chiral axis as $k$ traverses the Brillouin zone (see Fig.\ \ref{fig:ZakPhaseOfUtilde}c),
 and its Zak phase $\tilde\gamma$ (dashed line in Fig.\ \ref{fig:ZakPhaseOfUtilde}d) differs from the Zak phase $\gamma$ of protocol $U$ (dashed line in Fig.\ \ref{fig:ZakPhaseOfU}c). 

We realize experimentally protocol $\tilde{U}$ by the setup shown schematically in Fig.~\ref{fig:ZakPhaseOfUtilde}b. Using the relation $\sqrt{Q(\delta)}=Q(\delta/2)$, it is straightforward to see that $\tilde U^t=\sqrt{Q}WQW...QW\sqrt{Q}$. Hence, we realize the operator $\tilde U^t$ by placing $q$-plates yielding an optical retardation $\delta/2$ ($\sqrt{Q}$) at the beginning and end of the optical path, while in the bulk of the walk we adopt the same sequence reported in Fig.\ \ref{fig:ZakPhaseOfU}a (with the last $q$-plate removed). Overall, our quantum walk implements 7 steps of protocol $\tilde U$ by means of a total of eight $q$-plates, six with retardation $\delta$, two tuned at $\delta/2$ (first and last plates), separated by quarter-wave plates.
In Fig.\ \ref{fig:ZakPhaseOfUtilde}d, we report the measure of the mean chiral displacement $\tilde{\cd}$ generated by the single step operator $\tilde U$. As in the case of protocol $U$, this quantity accurately follows the theory prediction, providing an unambiguous detection of the Zak phase $\tilde\gamma$ of the infinite system after just 7 steps.

\begin{figure}[t!]
\centering
\includegraphics[scale=1]{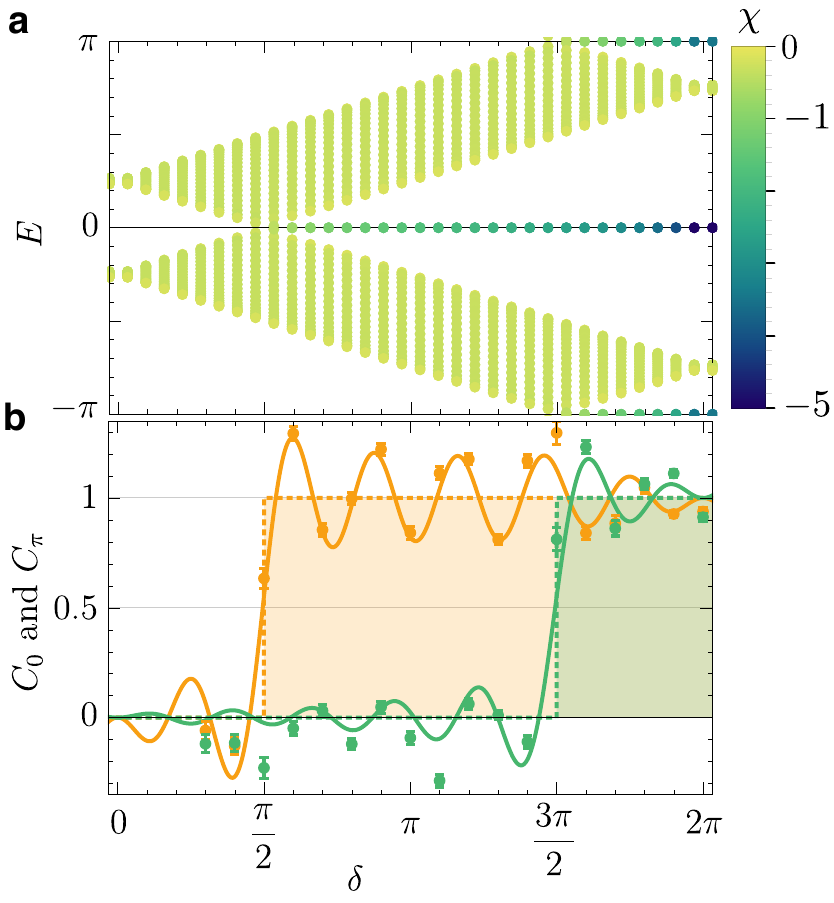}
\caption{{\bf Topological invariants and bulk-edge correspondence.} 
(a) Edge states on an open-ended lattice $[-L:L]$, with $L=10$; the color coding indicates the degree of localization $\chi=\log_{10}(1-\langle \vert m \vert\rangle/L)$, with darker colors indicating states more localized towards the edges. (b) Topological invariants $C_0$ and $C_\pi$, obtained as in equation \eqref{topInv} by combining the measurements of the mean chiral displacements $\cd$ and $\tilde\cd$ of protocols $U$ and $\tilde U$, and averaging the results obtained from the two initial states (error bars are the propagated standard errors).
The dashed lines show the long-time limit of the topological indices $C_0$ and $C_\pi$, yielding respectively the number of edge states at $0$- and $\pi$-energy.
}
\label{fig:edge_states_protocols_and_invariants}
\end{figure}

{\bf Complete topological characterization.}
It is clear from the previous discussion that the Zak phase associated with a single timeframe does not contain all the topological information of our QW. Indeed in Floquet 1D chiral systems there exist two independent classes of protected edge states at either $0$- and $\pi$-energies. An example of these edge states is shown in Fig.\ \ref{fig:edge_states_protocols_and_invariants}a, where we plot the quasi-energies of all eigenstates of an open-ended lattice. As remarked above, the spectrum is independent of the timeframe. 
The spectrum contains edge states even for $3\pi/2<\delta<5\pi/2$ where the Zak phase $\gamma$ of protocol $U$ is zero, explicitly confirming that the Zak phase of a single QW protocol does not contain the complete information about the topological state of the system.

The bulk-edge correspondence in these driven systems requires two invariants $C_0$ and $C_\pi$, yielding respectively the number of $0$- and $\pi$-energy edge states.
As shown in Refs.\ \cite{Asboth2013,Asboth2014}, these are simple functions of two Zak phases, measured in two inequivalent timeframes possessing an ``inversion point", i.e., which may be written respectively as $\mathcal{U}_1=\Gamma F^\dagger \Gamma F$ and $\mathcal{U}_2=F\Gamma F^\dagger \Gamma$, with $F$ a suitable evolution operator.
In the case of our setup, the two special protocols fulfilling this criterion are $\tilde U$ and $\tilde U'\equiv \sqrt W \cdot Q \cdot \sqrt W$. However it is simple to show that $\tilde U'$ is topologically equivalent to $U$, as no gap closing happens during the rotation $\sqrt{W}$; therefore the Zak phase of $\tilde U'$ coincides with $\gamma$. As such, the complete topological classification of 1D chiral systems may be obtained by means of the two quantities
\beq\label{topInv}
C_0= \frac{\tilde{\cd}+\cd}{2\pi} \quad {\rm and } \quad C_\pi= \frac{\tilde{\cd}-\cd}{2\pi},
\eeq
which converge in the long time limit, respectively, to the number of 0- and $\pi$-energy edge states. By combining our measurements of the mean chiral displacements measured in the inequivalent timeframes we are now able to compute the invariants $C_0$ and $C_\pi$ and detect the complete phase diagram of this system: the result is shown in Fig.\ \ref{fig:edge_states_protocols_and_invariants}b. Once again, our measurements show a remarkably fast convergence towards the asymptotic limit.

{\bf Robustness to dynamical disorder.}
Finally, we test the stability of the quantization of the mean chiral displacement against disorder. 
In particular, we choose protocol $U$, and introduce dynamical disorder by offsetting the optical retardation $\delta_j$ ($1\leq j\leq 7$) of 
  each  $q$-plate by a small random amount $|\epsilon_j|<\Delta$ around their common mean value $\bar\delta$. In our experiment, we set $\Delta=\pi/10$ and $\pi/5$.
We note that this disorder is dynamic, in the sense that it affects independently the various $q$-plates crossed by the beam, but crucially it respects chiral symmetry. This can be simply understood by noting that the vector $\vect{v}_\Gamma$, defining the chiral operator, does not depend on $\delta$.

\begin{figure*}[ht!]
\centering
\includegraphics[width=\linewidth]{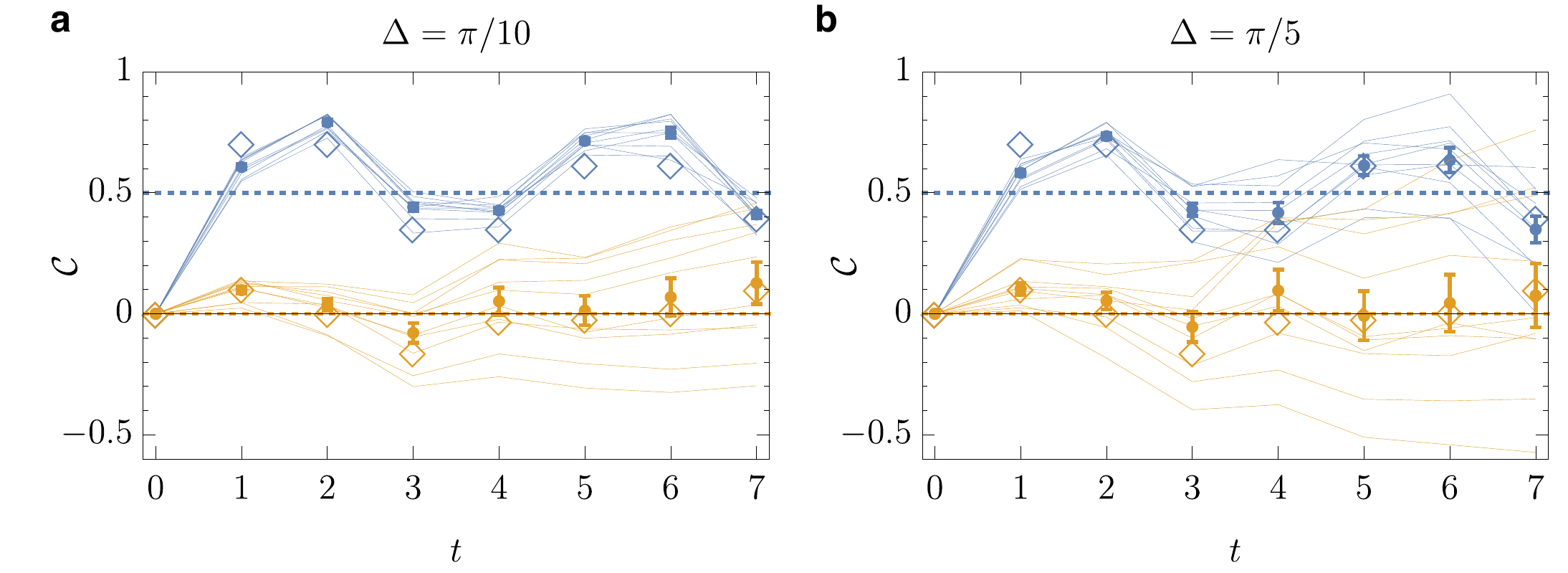}
\caption{{\bf Robustness to dynamical disorder.} Measurement of the mean chiral displacement $\cd$ of protocol $U$ for a localized input state in presence of dynamical disorder. For the orange (blue) lines, we choose a mean value of the q-plate optical retardation $\bar\delta = 7\pi/4$ ($\bar\delta=\pi$), expected to yield a Zak phases of $\gamma/2\pi=0$ ($\gamma/2\pi=1/2$), and we add at each time step a small random retardation $|\epsilon|<\Delta$, with $\Delta=\pi/10$ (left) and $\pi/5$ (right). Thin solid lines display the measurements of single realizations, and their average is shown as filled circles (error bars are the standard error of the mean). In all plots, empty diamonds represent theoretical simulation calculated for the ideal case $\Delta=0$, and dotted lines the expected result for $t\rightarrow\infty$.} 
\label{fig:disorder}
\end{figure*}

As shown in Fig.\ \ref{fig:disorder}, in single realizations the mean chiral displacement presents oscillations featuring higher amplitude for increasing disorder, but an ensemble average over independent realizations smoothly converges to the expected theoretical result, which in the infinite time limit gives the bulk value of the Zak phase. 
Here we performed measurements on protocol $U$, but similar robustness of the chiral displacement shall hold for every 1D QW chiral protocol, and more generally every 1D chiral system, as long of course as the disorder does not break chiral symmetry and its strength is small compared to the gap size to prevent inter-band transitions. 
As an example, in the Supplementary Note 2 and Supplementary Figs. 2-5 we show that the mean chiral displacement is an equally robust topological marker for a completely different and static (i.e., not driven) system, the celebrated SSH model.

\section*{Discussion}
Summarizing, here we proposed an efficient method to measure the Zak phase of a chiral system
by direct observation of its free bulk dynamics. 
In particular, we showed that information on the topological phase of the bulk is encoded in the mean chiral displacement, an oscillatory quantity that rapidly converges to the Zak phase, and is robust against (chiral-preserving) disorder in both space and time.

We experimentally verified our findings by performing the first measurement of the Zak phase of a chiral quantum walk.
The physical platform we chose is a photonic setup based on the orbital angular momentum of a light beam, where the mean chiral displacement corresponds to the relative shift of the two chiral polarization components. A precise readout of the Zak phase was obtained after only 7 quantum walk steps.
We further used the same method to measure the Zak phase in a complementary timeframe, which we realized by swapping few optical components. 
By combining the two measurements we extracted the two invariants providing the complete bulk-edge correspondence for this driven system, i.e., the one associated to the 0-energy edge state, and the one connected to the anomalous $\pi$-edge state.
Finally, we proved that the mean chiral displacement is a robust measure of the Zak phase by introducing dynamical but chiral-preserving disorder.

Although here we investigated experimentally a specific quantum walk, our results are not restricted to QWs, nor to Floquet systems. 
Indeed, the mean chiral displacement provides a robust topological characterization of arbitrary spin-1/2 1D chiral systems, either static or periodically-driven. These may nowadays be realized in a variety of platforms, ranging from ultracold atoms in optical lattices to photonic waveguides, and from semiconductor quantum wells to optomechanical systems. 

While formerly known methods for detection of topological properties require a uniform filling of the band of interest, external forces, loss mechanisms, or fine-tuning so that only edge states are populated, the method proposed here quite remarkably achieves this goal by observing the free evolution of a single particle, initially localized on a single site in the bulk.
This aspect may be specially beneficial for systems where filling a band is intrinsically challenging, such as bosonic condensates or phononic and photonic ensembles. 

Future interesting directions opened by this work include the extension of our results to chiral systems with more than two internal states, a further understanding of the role played by temporal disorder, and the topological characterization of systems in higher spatial dimensions.


\begin{footnotesize}

{\nolinenumbers \section*{Methods}}

{\bf Experimental setup.}
Our apparatus is shown schematically in Figs.\ \ref{fig:ZakPhaseOfU}a and \ref{fig:ZakPhaseOfUtilde}b, and a more detailed description is given in Supplementary Fig. 1. We produce a TEM$_{00}$ mode by coupling the output of a Ti:Sa laser ($\lambda=800$ nm) to a single mode fiber (SMF), thus preparing the beam in an orbital angular momentum (OAM) state with $m=0$. At the exit of the fiber, a specific polarization is selected by means of a sequence of a quarter-wave plate and a half-wave plate. Therefore the initial state of the QW is $\ket{\psi_0}=\ket{m=0}\otimes\ket{\bf s}$, where $m$ is the position in the walker (OAM) space \and ${\bf s}$ its coin state (polarization). 
In the standard protocol $U=Q\cdot W$ the single step consists of a quarter-wave plate oriented at 90$^\circ$ with respect to the horizontal direction (operator $W$), followed by a $q$-plate (operator $Q$), as shown in Fig.\ \ref{fig:ZakPhaseOfU}a. 
To implement the second protocol, we exploited the fact that equation \eqref{eq:qplate} may be written as $Q={\rm exp}(-i \delta \bn_Q\cdot\vectgr{\sigma})$, i.e., it corresponds to a rotation around a suitable unit vector $\bn_Q$; as such, a $q$-plate with retardation $\delta/2$ implements the desired operator $\sqrt Q$.
To implement the single step operator $\tilde U=\sqrt{Q} \cdot W \cdot \sqrt{Q}$ we then added a $\delta/2$ $q$-plate at the beginning of the sequence, and we halved the retardation of the last $q$-plate, as shown in Fig.\ref{fig:ZakPhaseOfUtilde}b.

{\bf $q$-plates.}
Each $q$-plate consists of a thin layer of birefringent liquid crystals, whose optic axes are arranged in a singular pattern characterized by a topological charge $q$ (in our case $q=1/2$). 
The patterned birefringence gives rise to an optical spin-orbit coupling that induces the polarization-dependent shift of OAM. 
Along with the specific pattern, the action of each device is determined by its optical retardation $\delta$, as reported in equation \eqref{eq:qplate}.
The optical retardation can be continuously tuned applying an electric field, allowing in turn for an accurate control of the spin-orbit interaction \cite{Piccirillo2010}.

{\bf Detection of the chiral displacement.} At the end of the walk we can select any polarization component of the final state by a combination of a quarter-wave and a half-wave plate, followed by a linear polarizer, and we measure its OAM content by diffraction on a spatial light modulator
coupled to a SMF and a power meter, which records the light intensity.
Since we are interested in analyzing the OAM spectrum of chiral components of the final wavepacket, waveplates orientations are selected so as to implement polarization projections onto chiral states $\chiru$ and $\chird$.
The chiral operators for protocols $U$ and $\tilde U$ are, respectively, $(\sigma_y+\sigma_z)/\sqrt2$ and $\sigma_z$, so it is straightforward to see that $\chiru_U=\cos{(\pi/8)}\ket{L}+i\,\sin{(\pi/8)}\ket{R}$ and $\chird_U=\sin{(\pi/8)}\ket{L}-i\,\cos{(\pi/8)}\ket{R}$ for protocol $U$, while $\chiru_{\tilde U}=\ket{L}$ and $\chird_{\tilde U}=\ket{R}$ in protocol $\tilde U$. 
The combination of polarization and OAM projections allows for determining the probabilities $P_{i,m}$, with $i=\{\uparrow,\downarrow\}$, that the system is in the chiral state $\ket{i}$ and in the OAM state $\ket{m}$. Given the probability distributions $P_{i,m}$, 
the chiral displacement is simply given by 
$\sum_m m \left(P_{\uparrow,m} - P_{\downarrow,m}\right)$.

{\bf Data availability.} The complete set of raw data supporting the findings of this study is available from the corresponding
author upon %
 request.

\end{footnotesize}
\vspace{10 mm}
\bibliography{TopQW}



{\nolinenumbers \section*{Acknowledgements}}
We acknowledge insightful discussions with J{\'a}nos Asb{\'{o}}th, Alessio Celi, and Miguel-Angel Martin Delgado. We thank Damiano Fiorillo for his help in taking and analyzing the experimental data. We acknowledge support from Adv.\ ERC grants PHOSPhOR and OSYRIS, EU grant QUIC (H2020-FETPROACT-2014 No.\ 641122), MINECO (Severo Ochoa grant SEV-2015-0522 and FOQUS FIS2013-46768), Generalitat de Catalunya (SGR 874 and CERCA), and the Fundaci\'o Privada Cellex. ADa acknowledges funding from the Cellex-ICFO-MPQ fellowship.
PM acknowledges funding from a ``Ram\'on y Cajal" fellowship.

\bigskip
{\nolinenumbers \section*{Author contributions statement}}
ADa, ML and PM, helped by FC, ADe, and MM, developed the complete theoretical framework,  building on preliminary studies of the dynamical subleading terms carried out by ES, GDF and VC.
FC, ADe, MM and LM designed the experimental methodology. 
FC and ADe, with contributions from MM and CdL, carried out the experiment. BP prepared the $q$-plates.
FC, ADe, ADa, MM, ML and PM wrote the manuscript. 
LM and ML supervised the project.
All authors discussed the results and contributed to refining the manuscript.

\bigskip
{\nolinenumbers \section*{Competing financial interest}}
The authors declare no competing financial interests.



\widetext
\clearpage



\setcounter{equation}{0}
\setcounter{figure}{0}
\setcounter{table}{0}
\setcounter{page}{1}
\makeatletter
\renewcommand{\figurename}{Supplementary Figure}

\clearpage

\section*{EXPERIMENTAL SETUP}

\begin{figure*}[ht!]
\centering
\includegraphics[width=0.8\linewidth]{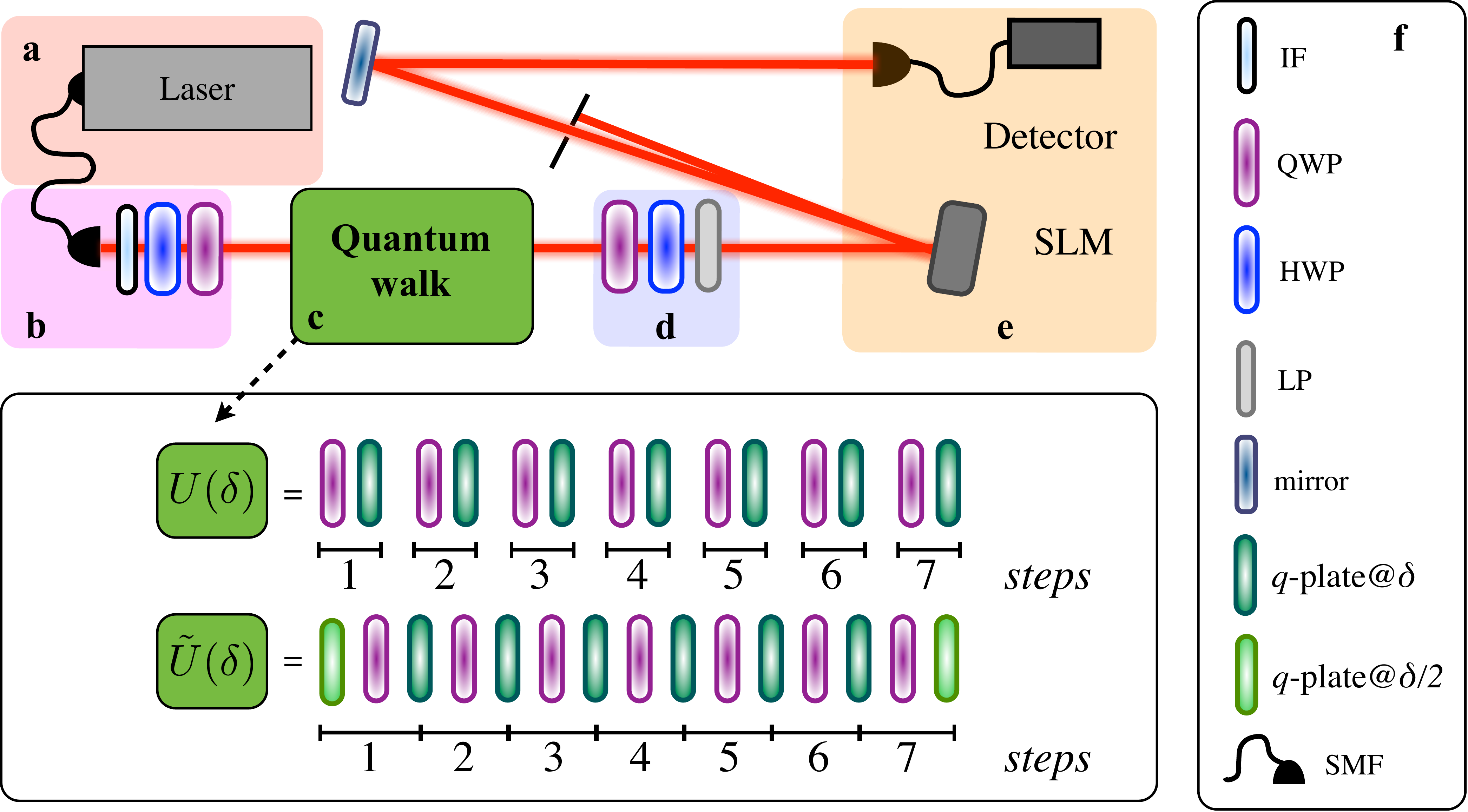}
\caption{{\bf Scheme of the experimental setup}. 
{\bf (a)} The output of a Ti:Sa pulsed laser source (pulse duration = 100 fs, central wavelength = 800 nm, repetition rate = 82 MHz) is coupled into a single mode fiber (SMF) so as to clean the laser spatial mode; this provides a single OAM state with $m=0$ at the input of the QW.  
{\bf (b)} At the exit of the fiber, the beam passes through an interferential filter (IF), whose transmittance is peaked at 800 nm with bandwidth of 3 nm, which allows to have a stable control of the light's wavelength and a narrower frequency distribution. 
Then the desired input polarization state is prepared by means of a half-wave plate (HWP) and a quarter-wave plate (QWP).  
{\bf (c)} The light beam passes through a sequence of QWPs and $q$-plates, as shown in detail in the inset, which are positioned in order to realize either protocol $U$ or $\tilde U$. 
{\bf (d)} At the end of the QW, a polarization component is selected by means of a QWP and a HWP, followed by a linear polarizer (LP). 
{\bf (e)} The OAM spectrum is measured by diffraction on a spatial light modulator (SLM), that displays standard pitchfork holograms for the projection over OAM states. At the first diffraction order, the light is coupled into a SMF that is directly connected to a power meter recording the field intensity. {\bf (f)} Legend of optical components displayed in panels {\bf (a}-{\bf e)}.
}
\label{fig:setup_SM}
\end{figure*}
\clearpage
\section*{Supplementary Note 1: Mean displacements}
We provide here the detailed derivation of the two key formulas introduced in the text, which yield the mean displacement of the whole wavepacket, and its mean chiral displacement.
To start, consider the evolution of a wavepacket $|\psi_0\rangle$ initially localized at site $m=0$, and let its polarization be characterized by  the expectation values of the three Pauli operators $\vect{s}=\langle \vectgr{\sigma} \rangle_{ \psi_0}\equiv\langle \psi_0 | \vectgr{\sigma} | \psi_0\rangle$.
The evolution operator generating $t$ timesteps of protocol $U$ is 
$U^t=(Q.W)^t={\rm e}^{-i E t \bn\cdot\vectgr{\sigma}}
=\cos(Et)\sigma_0 - i \sin(Et) \bn\cdot\vectgr{\sigma}$.
Using of the standard identity $({\bf a\cdot\vectgr{\sigma}})({\bf b}\cdot\vectgr{\sigma})=({\bf a \cdot \bf b})\sigma_0+i({\bf a \times b})\vectgr{\sigma}$ (valid for arbitrary vectors ${\bf a}$ and ${\bf b}$), it is straightforward to show that the mean displacement of the wavepacket reads:
\begin{equation}\label{meanDisplacementFullCalculation}
\begin{split}
\langle m(t)\rangle&= \bra{\psi(t)} m \ket{\psi(t)} = 
 \int_{-\pi}^{\pi}\frac{{\rm d}k}{2\pi}\left\langle U^{-t} 
(i \partial_k)
 U^t \right\rangle_{\psi_0}\\
&= \int_{-\pi}^{\pi}\frac{{\rm d}k}{2\pi} \langle \left[\cos(Et)\sigma_0+ i \sin(Et)\bn\cdot \vectgr{\sigma}\right] 
(i \partial_k)
\left[\cos(Et)\sigma_0- i \sin(Et)\bn\cdot\vectgr{\sigma}\right] \rangle_{\psi_0}\\
&= 
{
\int_{-\pi}^{\pi}\frac{{\rm d}k}{2\pi} \left[\left(t\frac{\partial E}{\partial k}
-\frac{\partial}{\partial k}\frac{\sin(2Et)}{2}\right) \vect{n}
-\sin(Et)^2\left(\vect{n}\times\frac{\partial \vect{n}}{\partial k}\right)\right]\cdot\vect{s}\;.
}
\end{split}
\end{equation}
When a protocol possesses chiral symmetry (and both ours do), the unit vector $\bn$ of the corresponding effective Hamiltonian is bound to rotate in the plane orthogonal to the vector ${\bf v}_\Gamma={\rm tr}(\Gamma\vectgr{\sigma})/2$ associated to its chiral operator $\Gamma$. In turn, this means that $(\bn\times\partial_k\bn)$ is parallel to ${\bf v}_\Gamma$. We arrive then to the expression quoted in the text,
\beq\label{finalExprForMeanDispl}
\langle m(t)\rangle=\langle \Gamma_\perp \rangle_{ \psi_0} [L(t)+S(t)]{-}\langle \Gamma \rangle_{ \psi_0} S_\Gamma(t).
\eeq
Specifically, the chiral operator for protocol $U$ is $\frac{\sigma_y+\sigma_z}{\sqrt{2}}$, so that  $\langle \Gamma \rangle_{ \psi_0}=\frac{s_y+s_z}{\sqrt{2}}$, and one finds $\langle \Gamma_\perp \rangle_{ \psi_0}=\frac{s_y-s_z}{\sqrt{2}}$.

We omit here for simplicity the explicit expressions of the functions $\{E,\,\bn,\, L,\,S\}$ in terms of the optical retardation $\delta$ and the quasi-momentum $k$, as these are protocol-dependent, rather bulky, and not particularly illuminating. However, it is obvious that $E$ and $\bn$ are independent of time, so the term $L(t)$ in Supplementary Eq.\ \eqref{finalExprForMeanDispl} is ballistic (i.e., grows linearly with $t$), while $S(t)$ is oscillatory. The last function in Supplementary Eq.\ \eqref{finalExprForMeanDispl} is the one we refer to as {\it chiral term},
\beq\label{chiralTerm}
S_\Gamma(t)=\int_{-\pi}^{\pi}\frac{{\rm d}k}{2\pi} \sin(Et)^2
\left(\vect{n}\times\frac{\partial \vect{n}}{\partial k}\right)\cdot{\vect v}_\Gamma
=\frac{\gamma}{2\pi}-\int_{-\pi}^{\pi}\frac{{\rm d}k}{2\pi} \frac{\cos(2Et)}{2}
\left(\vect{n}\times\frac{\partial \vect{n}}{\partial k}\right)\cdot{\vect v}_\Gamma,
\eeq
which is clearly proportional to the Zak phase 
$\gamma={1 \over 2}\int_{-\pi}^{\pi}{\rm d}k(\vect{n}\times\partial_k\vect{n})\cdot{\vect v}_\Gamma$,
 plus an oscillatory contribution whose amplitude and period generally decay rapidly as $t\rightarrow\infty$.
A singular point is however $\delta=2\pi$, where $E=3\pi/4$, independent of $k$, so that $\cos(2Et)$ equals 0 for odd $t$, and $(-1)^{t/2}$ for even $t$. This means that $S_\Gamma(t)$ at $\delta=2\pi$ equals exactly $\gamma/2\pi$ for odd values of $t$, while it equals 0 ($\gamma/\pi$) for even (odd) values of $t/2$. For this reason, in Figs.\ 1-3 of the main text  we have considered quantum walks with an odd number of steps.

The derivation of the mean displacement for protocol $\tilde{U}=\sqrt{Q}\cdot W\cdot \sqrt{Q}$ yields formulas which are identical to Supplementary Eqs.\ \eqref{meanDisplacementFullCalculation}-\eqref{chiralTerm}, provided one replaces $\bn$ with $\tilde{\bn}$ and remembers that the chiral operator is $\sigma_z$, so that $\langle \tilde\Gamma \rangle_{ \psi_0}=s_z$, and $\langle \tilde\Gamma_\perp \rangle_{ \psi_0}=s_y$.

The mean chiral displacement $\cd$, Eq.\ \eqref{eq:meanChiralDisplacement} of the main text, may be computed along the same lines. We show here the calculation for the simplest case of the chiral displacement $\tilde{\cd}$ associated to the protocol $\tilde{U}$, whose unit vector $\tilde{\bn}$ has a vanishing $z$-component since $\tilde\Gamma=\sigma_z$. One finds:
\begin{equation}\label{derivchiral}
\begin{split}
\tilde\cd(t) &= {\langle \tilde\Gamma m(t)\rangle = 
\int_{-\pi}^{\pi}\frac{{\rm d}k}{2\pi} \left\langle \tilde{U}^{-t}\sigma_z (i \partial_k) \tilde{U}^t \right\rangle_{\psi_0}
}
\\
&=\int_{-\pi}^{\pi}\frac{{\rm d}k}{2\pi} \left\langle \sin(Et)^2\left[
\left(\tilde{\bn}\times\partial_k \tilde{\bn}\right)_z\sigma_0 -\frac{i}{2} \partial_k|\tilde{\bn}|^2 \sigma_z\right]+ 
i\partial_k\Big[\cos(Et)\sin(Et)(\tilde{n}_x\sigma_y-\tilde{n}_y\sigma_x)-\sin(Et)^2\sigma_z\Big]
\right\rangle_{\psi_0}\\
&=\int_{-\pi}^{\pi}\frac{{\rm d}k}{2\pi} \sin(Et)^2\left(\tilde{\bn}\times\frac{\partial \tilde{\bn}}{\partial k}\right)_z=S_{\tilde\Gamma}(t).
\end{split}
\end{equation}
In the second line of Supplementary Eq.\ \eqref{derivchiral}, all 
terms preceded by an imaginary unit $i$ integrate to zero: the first because $\tilde\bn$ is a vector of unit norm for all $k$, and the second because it is the integral of a total derivative over a closed path. The final result is purely real, in agreement with the fact that the chiral displacement is the expectation value of an Hermitian operator.
The derivation of the chiral displacement $\cd$ for protocol $U=Q\cdot W$ is completely analogue, provided the reference frame of the Pauli matrices is suitably rotated.

\section*{Supplementary Note 2: SSH model}

\begin{figure*}[h]
\centering
\includegraphics[scale=1]{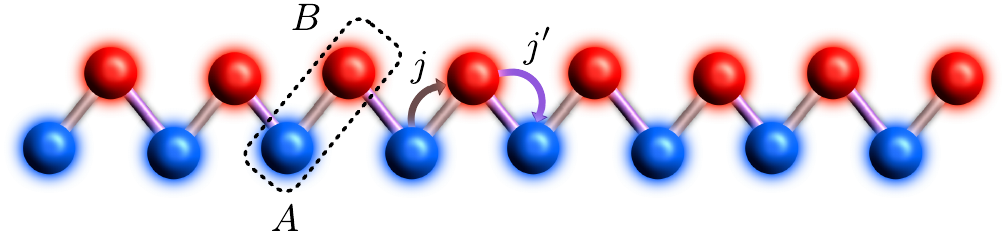}
\caption{{\bf SSH model.} The SSH model describes a one-dimensional lattice with each unit cell containing two sites, $A$ (blue) and $B$ (red). The Hamiltonian has an intra-cell hopping of amplitude $j$ and an inter-cell hopping of amplitude $j'$.}
\label{fig:SSH}
\end{figure*}

The Su-Schrieffer-Heeger (SSH) model, originally introduced to study polyacetylene chains, is probably the simplest and most studied chiral-symmetric topological system. It describes a one-dimensional dimerized lattice, where each unit cell contains two sites labeled $A$ and $B$. Further on, we will call them alternatively lattice sub-sites or  polarizations. The amplitudes $j$ and $j'$ quantify, respectively, the hopping between adjacent sub-sites belonging to the same cell, and to neighboring ones. The Hamiltonian describing the dynamics along a dimerized lattice with $N$ cells reads:
 \beq\label{eq:hamiltonian_ssh}
 H = j\sum_{n=1}^{N}  \bc^{\dagger}_n \sigma_x \bc_n + j'\sum_{n=1}^{N-1} \left(\bc^{\dagger} _{n+1} \sigma_+ \bc_{n} + \bc^\dagger_{n} \sigma_- \bc_{n+1} \right)
 \eeq
 Here $\mathbf{\bc}^\dagger_n =(c^\dagger_{n,A},c^\dagger_{n,B})^T$ creates a particle on site $A$ and $B$ of cell $n$, and $\sigma_i$ are Pauli matrices acting in the sub-lattice space.
 The Bloch Hamiltonian is a $2\times2$ matrix:
\beq\label{eq:bloch_SSH}
  H(k)= [j + j' \cos{(ka)} ]\,\hat{\sigma}_x + j' \sin{(ka)}\,\hat{\sigma}_{y}=E(k)\bn(k)\cdot\vectgr{\sigma},
\eeq
where the dispersion reads $E(k)=\pm\sqrt{j^2+j'^2+2jj'\cos(k)}$, the real unit vector $\bn(k)$ determines the position of the eigenstates on the Bloch sphere, and $a$ is the lattice constant.
Since the Hamiltonian doesn't contain direct couplings between $A - A$ or $B - B$ sites, the vector $\bn(k)$ lies in the $xy$ plane, so that the chiral symmetry operator $\Gamma$ is $\sigma_z$. Hamiltonians with $j>j'$  and $j<j'$ and are topologically inequivalent, being the winding number (i.e., the Zak phase $\gamma$ divided by $\pi$) equal to zero and one, respectively.

Equations \eqref{meanDisplacement} and \eqref{eq:meanChiralDisplacement} of the main text, giving respectively the mean displacement and the mean chiral displacement of an initially localized walker performing a chiral quantum walk, describe equally well the dynamics of a particle on the SSH chain, initially localized on a single unit cell. $A$ and $B$ sites of the SSH lattice may be identified with the $|L\rangle$ and $|R\rangle$ polarization states of the quantum walker.  

\begin{figure*}[h!]
\centering
\includegraphics[width=\linewidth]{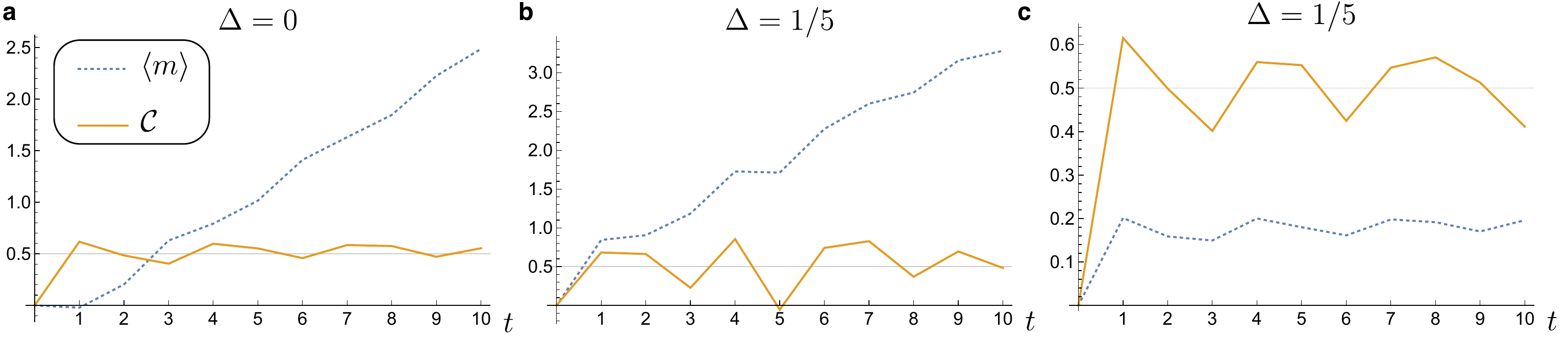}
\caption{{\bf Disordered SSH model.} Evolution of a walker on an SSH lattice with dynamical disorder in the tunnelings. At $t=0$, the walker is initialized on the central unit cell of the chain, with a random polarization (different for each realization), and we have taken $j=j'/2$, so that the model has a Zak phase $\gamma=\pi$. Dashed (solid) lines depict the mean (mean chiral) displacement. 
{\bf (a)} Single realization in absence of disorder. 
{\bf (b)} Single realization with disorder amplitude $\Delta=1/5$. 
{\bf (c)} Upon ensemble-averaging over 100 realizations (with $\Delta=1/5$), the mean chiral displacement smoothly converges to $\gamma/(2\pi)$.
}
\label{fig:disorderedSSH}
\end{figure*}

Exactly as for the QW analyzed in the main text, the mean chiral displacement is a robust marker of the topological phase of the SSH lattice. 
To illustrate this, we simulate the evolution of a localized electron in a lattice with $j'/j>1$, so that the Zak phase $\gamma$ is $\pi$.
As we show in panel (a) of Supplementary Fig.\ \ref{fig:disorderedSSH}, in absence of disorder the mean value of the chiral displacement oscillates around $\gamma/(2\pi)=1/2$, and converges to this value in the long time limit. 
To test the robustness of the measurement, we add disorder to the system, in the form of a random dynamical disturbance to each tunneling; specifically, at each timestep $t$ the $A-B$ tunneling in cell $n$, $j_{n,t}$, is replaced by $j_{n,t}+\epsilon_{n,t}$, with a randomly chosen $|\epsilon_{n,t}|<\Delta$ (and similarly for the tunneling $j'_{n,t}$ connecting cells $n$ and $n+1$); note that this form of the disorder explicitly preserves the chiral nature of the model. In presence of disorder, like in the QW case discussed in Fig.\ \ref{fig:disorder} of the main text, single realizations present oscillations of increasing amplitude with increasing disorder, but an ensemble average over independent realizations smoothly converges to the expected theoretical result; see panels (b) and (c) of Supplementary Fig.\ \ref{fig:disorderedSSH}.\\

\begin{figure*}[h!]
\centering
\includegraphics[scale=1]{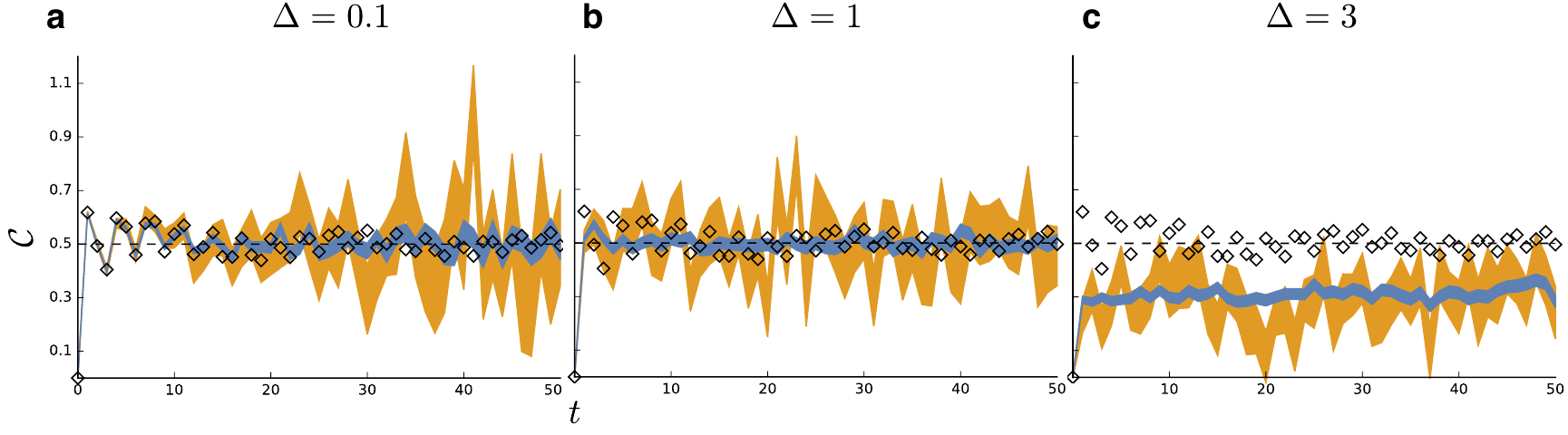}
\caption{\label{fig:SSH_disorder_1}
{\bf Long-time behavior with static disorder.} Mean chiral displacement $\mathcal{C}$ of the SSH model with $j'=2j$ and static disorder over all tunnelings, of amplitude \textbf{(a)} $\Delta=0.1$,\textbf{(b)} $\Delta=1$ and \textbf{(c)} $\Delta=3$. The yellow (blue) line indicates the ensemble average of $\mathcal{C}$ over 50 (1000) realizations, and the width of the line corresponds to the standard error of the mean. The black diamonds show the result in absence of disorder, where the spectrum has a gap of amplitude $\Delta_{\rm gap}=2$. The images show how, in the long time limit, the ensemble average of the mean chiral displacement remains locked to a value consistent with the Zak phase, as long as $\Delta\lesssim \Delta_{\rm gap}$.}

\end{figure*}

{\bf Behavior at very long times.} We focus here on the effect of disorder after very long times. 
In Supplementary Figure \ref{fig:SSH_disorder_1} we consider the mean chiral displacement $\mathcal{C}$ of the disordered SSH model with $j'=2j$. The yellow (blue) line displays the ensemble average over 50 (1000) realisations of static disorder with amplitude $\Delta$, while the black diamonds depict the mean chiral displacement in absence of disorder. 
In the clean case, the system presents an energy gap $\Delta_{\rm gap}=2$, 
and each energy band has a bandwidth of $\Delta_{\rm bw}=2$.
Supplementary Figure \ref{fig:SSH_disorder_1}a shows the effect for $\Delta=0.1$: the ensemble average converges to the mean chiral displacement, even in the long time limit.

\begin{figure*}[h!]
\centering
\includegraphics[scale=1]{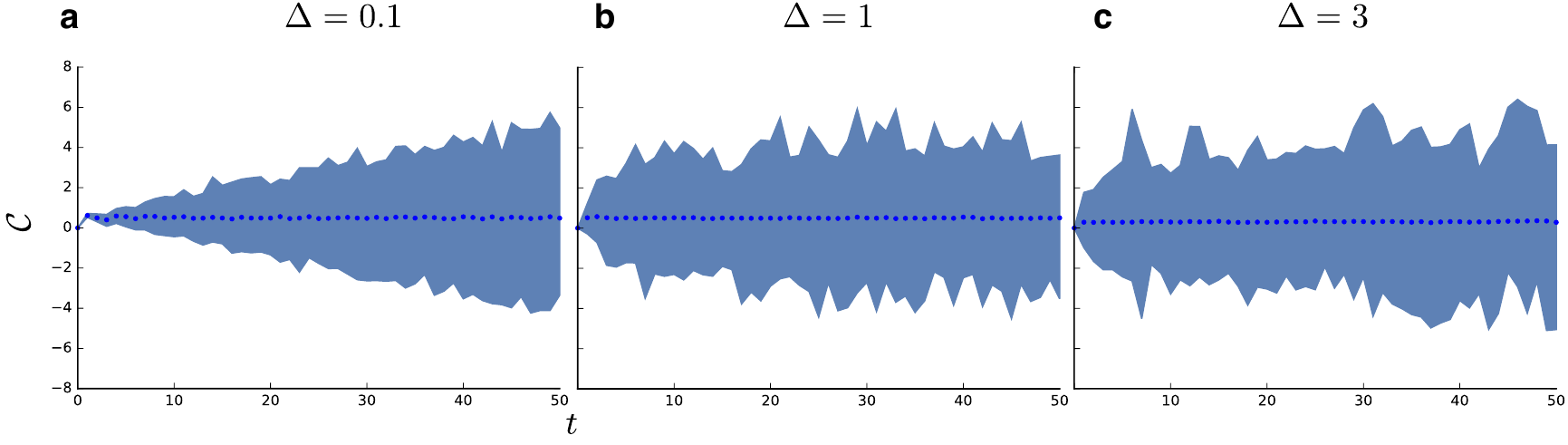}
\caption{\label{fig:SSH_disorder_2}
{\bf Effect of the disorder on the error in time.} The mean chiral displacement is plotted for each disorder realization (blue area) and is compared to the ensemble average (dots) for \textbf{(a)} $\Delta=0.1$,\textbf{(b)} $\Delta=1$ and \textbf{(c)} $\Delta=3$. For small disorder, the error on the ensemble average grows in time. For important disorder, the error on the ensemble average is independent of the time.}
\end{figure*}

Supplementary Figure \ref{fig:SSH_disorder_1}b shows the case of $\Delta=1$, a disorder smaller than the energy gap but comparable to the bandwidth.
Strikingly, in this case the ensemble averaged mean chiral displacement converges to the Zak phase faster than in absence of disorder (as may be noticed by comparing the blue line with the black diamonds).
This demonstrates that a moderate disorder ($\Delta\lesssim \Delta_{\rm gap}$) can wash out the coherent oscillations, but it does not destroy topological effects, and actually may even be beneficial, as it provides faster convergence: moderate disorder favors the measure of the topology. 
Supplementary Figure \ref{fig:SSH_disorder_1}c shows instead the result for a disorder sensibly larger than the energy gap, $\Delta=3$. In this case, as expected, topological protection is completely lost, and the ensemble averaged mean chiral displacement fails to converge to the Zak phase.\\ We continue the study of static disorder by commenting on the growth in time of the error on the ensemble average. 
In Supplementary Figure \ref{fig:SSH_disorder_2}, the blue area denotes the superposition of all 1000 values of $\mathcal{C}$ which were computed to produce Supplementary Figure\ \ref{fig:SSH_disorder_1}, and the bright blue dots denote their ensemble average.
For disorder amplitudes smaller than the bandwidth, we observe that $\mathcal{C}$ undergoes oscillations of steadily growing amplitude in time.  
Therefore, the longer the time of observation, the more realizations will be needed to minimize the error. 
When the amplitude of the disorder becomes comparable to the bandwidth, however, $\mathcal{C}$ shows large oscillations even at very short times, but their amplitude does not grow further in time.
Therefore, in the long time limit it is not necessary to increase the number of realisations to measure the Zak phase in presence of moderate disorder.

 \end{document}